\documentclass[aps,prl,twocolumn,groupedaddress,floatfix,superscriptaddress]{revtex4}
\usepackage{amssymb,amsmath}
\usepackage{graphicx}
\usepackage{subfigure}
\usepackage[english]{babel}
\usepackage{float}
\usepackage[edtable]{lineno}
\usepackage{soul}
\usepackage[usenames]{color}
\usepackage{relsize}
\usepackage{printlen}

\usepackage{color}
\definecolor{SMblack}{RGB}{00,00,00}
\definecolor{Seba}{RGB}{00,00,00}
\usepackage[colorlinks=true,
            linkcolor=magenta,
            urlcolor=blue,
            citecolor=blue]{hyperref}

\usepackage{footmisc}
\DefineFNsymbols{mySymbols}{{\ensuremath\dagger}{\ensuremath\ddagger}\S\P
   *{**}{\ensuremath{\dagger\dagger}}{\ensuremath{\ddagger\ddagger}}}
\setfnsymbol{mySymbols}

\begin{document}

\title{Observation of Ground and Excited Flat Band States in Graphene Photonic Ribbons}

\author{Camilo Cantillano\footnote{These authors contributed equally to this work.}}
\affiliation{Departamento de F\'{\i}sica and MSI-Nucleus on Advanced Optics, Facultad de Ciencias, Universidad de Chile, Santiago, Chile}
\author{Sebabrata Mukherjee$^{\dagger,}$}
\email[Email: ]{s.mukherjee@hw.ac.uk}
\affiliation{Scottish Universities Physics Alliance (SUPA), Institute of Photonics and Quantum Sciences, School of Engineering $\&$ Physical Sciences, Heriot-Watt University, Edinburgh, EH14 4AS, United Kingdom}
\author{Luis Morales-Inostroza}
\affiliation{Departamento de F\'{\i}sica and MSI-Nucleus on Advanced Optics, Facultad de Ciencias, Universidad de Chile, Santiago, Chile}
\author{Basti\'an Real}
\affiliation{Departamento de F\'{\i}sica and MSI-Nucleus on Advanced Optics, Facultad de Ciencias, Universidad de Chile, Santiago, Chile}
\author{Gabriel C\'aceres-Aravena}
\affiliation{Departamento de F\'{\i}sica and MSI-Nucleus on Advanced Optics, Facultad de Ciencias, Universidad de Chile, Santiago, Chile}
\author{Carla Hermann-Avigliano} 
\affiliation{Departamento de F\'{\i}sica and MSI-Nucleus on Advanced Optics, Facultad de Ciencias, Universidad de Chile, Santiago, Chile}
\author{Robert R.~Thomson}
\affiliation{Scottish Universities Physics Alliance (SUPA), Institute of Photonics and Quantum Sciences, School of Engineering $\&$ Physical Sciences, Heriot-Watt University, Edinburgh, EH14 4AS, United Kingdom}
\author{Rodrigo A.~Vicencio}
\email[Email: ]{rvicencio@uchile.cl}
\affiliation{Departamento de F\'{\i}sica and MSI-Nucleus on Advanced Optics, Facultad de Ciencias, Universidad de Chile, Santiago, Chile}



\begin{abstract}
Understanding the wave transport and localisation is a major goal in the study of lattices of different nature. In general, inhibiting the energy transport on a perfectly periodic and disorder-free system is  challenging, however, some specific lattice geometries allow localisation due to the presence of dispersionless (flat) bands in the energy spectrum.  
Here, we report on the experimental realisation of a quasi-one-dimensional photonic graphene ribbon supporting four flat-bands. We study the dynamics of fundamental and dipolar modes, which are analogous to the $s$ and $p$ orbitals, respectively. 
In the experiment, both modes (orbitals) are effectively decoupled from each other, implying two sets of six bands, where two of them are completely flat. 
Using an image generator setup, we excite the $s$ and $p$ flat band modes and demonstrate their non-diffracting propagation for the first time. Our results open an exciting route towards photonic emulation of higher orbital dynamics.
\end{abstract}

\maketitle

The physics of graphene is an intense field of research primarily focused on their unique electronics and magnetic properties. In particular, graphene nanoribbons can exhibit edge states~\cite{gr1} as well as the transition from semiconductors to semi-metals, depending on the number of coupled ribbons~\cite{gr2,gr3}. Several attempts to fabricate and characterize these graphene-like structures have been reported due to their fundamental relevance for future applications in nanoelectronics~\cite{gr4,gr5}. This includes room-temperature ballistic transport~\cite{gr6}, well-controlled atomic configurations~\cite{gr7}, photonics and optoelectronic applications~\cite{gr8}. In the photonic platform, graphene lattices have already been induced in photorefractive crystals at 
{\color{SMblack} the micrometer} scale, where conical diffraction and nonlinear localisation were experimentally observed~\cite{hcmoti}. Additionally, the observation of unconventional edge states~\cite{hcedge}, photonic floquet topological insulators~\cite{htopo}, and pseudospin-mediated vortex generation~\cite{hcpseudo} have been reported in graphene optical lattices. The ability of directly imaging the wavefunction gives an important experimental advantage for photonic setups~\cite{rep1, rep2, Garanovich2012}, in comparison to 
solid-state physics.

Recent advancement in experimental physics enabled us to emulate various semi-classical and quantum phenomena in a highly controllable environment. Ultracold atoms in optical lattices~\cite{Bloch2005, Jaksch2005} and periodic arrays of coupled optical waveguides (photonic lattices)~\cite{rep1,rep2, Garanovich2012} are two parallel experimental platforms which were extensively used to observe and probe various intriguing solid-state phenomena. This includes the localisation effects induced by external fields~\cite{Dreisow2008, WSL}, disorder~\cite{Schwartz2007, Billy2008} and particle interactions~\cite{Greiner2002, Szameit2006}. Indeed, localisation is a major goal in diverse areas of physics, where the trapping and control  of excitations of different nature become crucial~\cite{rep2}. During several years, photonics has taken a central role on this problem, being particularly intense in the context of photonic lattices. Different fabrication techniques have been developed, being the femtosecond-laser technique probably the most flexible one in order to fabricate arbitrary three-dimensional configurations~\cite{fslt,femtoseba}. Most of the known methods to localise energy rely on the modification of the lattice using linear or nonlinear defects, or by destroying the periodicity of the system. However, localised states in a photonic Lieb lattice~\cite{lieb2,liebseba} were recently observed in the linear optical regime, due to the existence of a completely flat-band (FB). The states residing on this non-diffractive band occupy only a few sites and can be considered as localised states in the continuum~\cite{bic,lieb2}.
Unfortunately, the flatness of a given band can be modified if extra interactions are also considered in the model~\cite{desa1}. This is a frequent problem on several FB systems 
which diminishes the chances for an experimental excitation of FB localised states. However, by inspecting the discrete properties of a given system, it is possible to identify some lattices where next or even next-next nearest neighbour (NN) interactions preserve the flatness of the band. This requires a high degree of symmetry in order to effectively cancel the transport at different connector sites~\cite{luis}.

Almost all the experimental research devoted to the study of periodical systems has focused on the excitation of fundamental modes on different lattice sites. This is essentially due to experimental complications of exciting higher order modes, which in the case of cold atoms have been solved indirectly by selectively populating p-band states~\cite{becdipole}. However, a precise excitation of dipolar states has only been possible very recently on optical waveguide lattices using an image generator setup~\cite{dipole1}, where a {\color{SMblack}well-defined} contrast between the transport of fundamental and dipolar states has been shown. The possibility to experimentally excite and control higher bands excitations, in optical lattice systems, paves the venue in which the study of remarkable properties of correlated systems such as superfluidity, superconductivity, antiferromagnetic ordering, among others, becomes possible~\cite{wu, li1, yin, li2}.

In this Article, we study theoretically and experimentally a graphene-like ribbon {\color{SMblack} where each lattice site supports two non-degenerate modes, the fundamental and dipolar modes.} This system is particularly interesting because it {\color{SMblack} can possess} 
two flat bands per mode, and these bands are robust against higher-order coupling interactions. This implies that the excitation of FB states is quite stable in realistic experimental conditions, as we show below. 
As these modes possess a large propagation constant detuning, the interaction between them is effectively absent in the dynamics. 
To the best of our knowledge, this is the first experimental 
{\color{SMblack} realisation} of a periodical lattice possessing 
{\color{SMblack} multiple flat-bands,} corroborated by the 
{\color{SMblack}observation of the spatially localised flat-band states.} 

\begin{figure}[t!]
\centering
\includegraphics[width=0.9\linewidth]{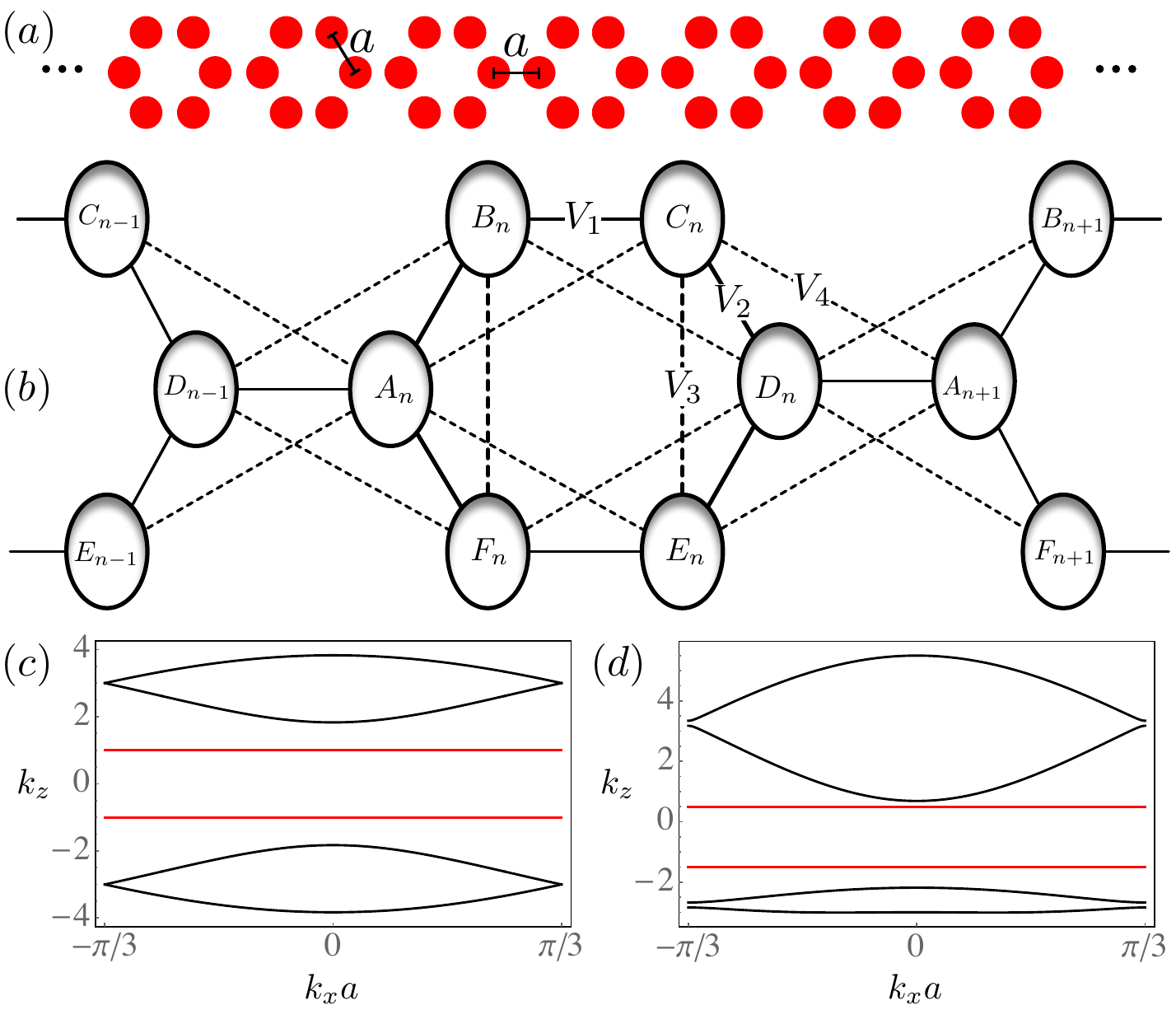}
\caption{{\bf } (a) A graphene-ribbon lattice. (b) Couplings interactions represented by lines. Linear spectrum for (c) $V_1=1$, $V_2=2$, $V_3=V_4=0$ and (d) $V_1=1$, $V_2=2$, $V_3=V_4=0.5$.
\label{f1}}
\end{figure}

{\it Model.}    
The unit cell of a graphene ribbon consists of a sequence of six-sites as sketched in Fig.~\ref{f1}(a), where each waveguide is separated from its nearest neighbour by a centre-to-centre distance ``$a$''. The interaction between lattices sites is governed by the evanescent coupling which decreases exponentially with the distance between waveguides~\cite{fslt,femtoseba}. We define the nearest and next-nearest neighbour coupling coefficients in Fig.~\ref{f1}(b), where the horizontal coupling is $V_1$, the short-diagonal one is $V_2$, the vertical coefficient is $V_3$, and the long-diagonal one is $V_4$. The contribution of all other long range couplings can be safely neglected for the maximum propagation distance considered here. For our laser inscribed photonic lattice (PL), each waveguide supports elliptically oriented modes with the major axis along the vertical, implying that $V_2>V_1$ and $V_3>V_4$.

In the scalar-paraxial approximation, the evolution of light waves across a graphene ribbon is governed by the following discrete linear Schr\"odinger-like equations~\cite{rep1, rep2, Garanovich2012}
\begin{equation}
-i\frac{\partial\psi_{\vec n}^j}{\partial z}=\beta_j\psi_{\vec n}^j+\sum_{\vec m\neq\vec n}V_{\vec n,\vec m}^j\psi_{\vec m}^j\ .
\label{eq1}
\end{equation}
Here, $\psi_{\vec n}^j$ describes the field amplitude of a given mode, $j\!=\!\{s, p$\}, at the $\vec n$-th site, with propagation constant $\beta_j$, $z$ corresponds to the propagation coordinate (dynamical variable) along the waveguides, and $V_{\vec n,\vec m}^j$ represents the coupling interactions between sites $\vec n$ and $\vec m$ for mode $j$. In model (\ref{eq1}), it was assumed that the $s$ and $p$ modes are effectively decoupled. First, we consider that each waveguide supports only a single mode, for {\color{SMblack}example,} the fundamental $s$ mode. In order to find the linear spectrum of this lattice, we first define the unitary cell composed of sites $A$, $B$, $C$, $D$, $E$, and $F$ as shown in Fig.~\ref{f1}(b), and insert a plane wave ansatz $\vec{\Psi}_n(z)=\vec{\Psi}_0\exp(i k_x a n) \exp(i k_z z)$, with $\vec{\Psi}_l\equiv \{A_l,B_l,C_l,D_l,E_l,F_l\}$. Here, $k_x$ and $k_z$ {\color{SMblack}correspond} to the transverse and longitudinal propagation constants, 
respectively. By solving the eigenvalue problem, we identify two flat bands, $k_z^{\pm}(k_x)=\pm V_1-V_3$, with degenerate eigenmodes, as indicated by the red horizontal lines in Figs.~\ref{f1}(c) and (d). It should be highlighted that the flatness of these two bands is independent of the next-nearest neighbour interactions due to the symmetry of the lattice geometry. 
Only for a reduced set of parameters, the rest four linear bands can be expressed in a closed form. Therefore, for generality, we show the band structure in Fig.~\ref{f1} for two different cases considering (c) only NN and (d) NN plus next NN interactions. In both cases, one can observe two perfectly flat bands, demonstrating the robustness of FB phenomena against the next-nearest neighbour interactions in this lattice geometry. The compact localised states occupy only four sites ($B$, $C$, $E$, and $F$) of a unit cell, with equal intensity and the following phase distributions: 
{\color{SMblack} $\{+,+,-,-\}$} for the upper and 
{\color{SMblack} $\{+,-,+,-\}$} lower flat bands, {\color{SMblack}respectively}. We can easily identify the destructive interference at sites $A_n$ and $D_n$, as expected considering the properties of mini-arrays~\cite{luis}. When exciting these localised FB states, the transport is absolutely cancelled across the lattice due to the perfectly zero amplitude at the connector sites.

\begin{figure}[t!]
\centering
\includegraphics[width=0.95\linewidth]{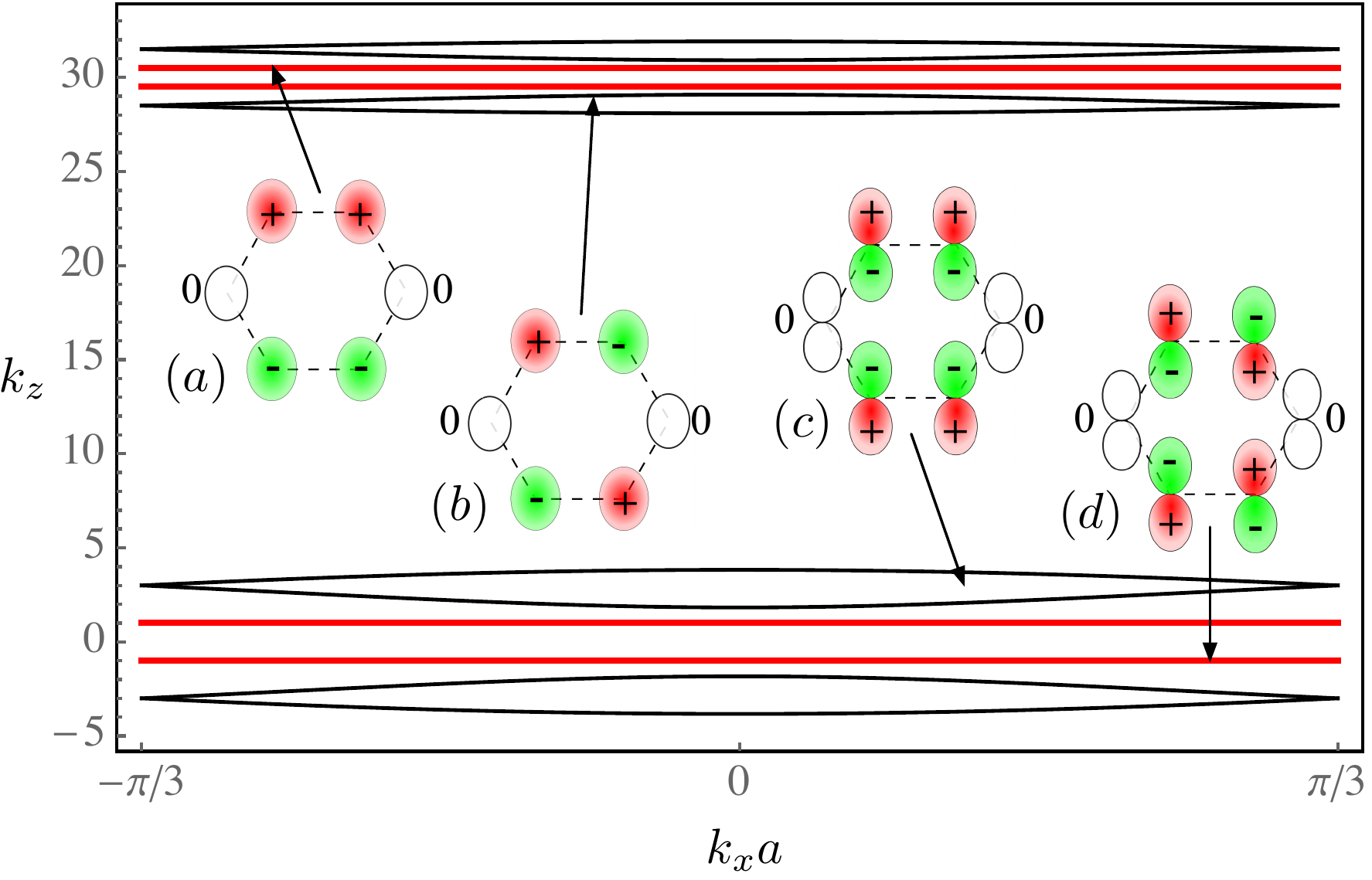}
\caption{{\bf } Composed linear spectrum of a  graphene-ribbon with $V_1^s\!=\!0.5$, $V_2^s\!=\!1$ and $V_1^p\!=\!1$, $V_2^p\!=\!-2$. (a)--(d) Intensity and phase profiles of the FB modes. Here, red (green) colour represents a positive (negative) phase.
\label{f2}}
\end{figure}

Now, we consider that each waveguide in the lattice supports two modes, the fundamental  ($s$) and the vertically-oriented dipolar ($p$) modes. It should be mentioned that the propagation constant (or the analogous energy) of the supported modes, and hence, the excitation of {\color{SMblack}higher-order} modes, can be efficiently controlled by tuning the wavelength $\lambda$ of incident light. The coupling between the two modes at the same lattice site is forbidden due to orthogonality. The large mismatch in propagation constants (defined as $\Delta \beta\equiv |\beta_s-\beta_p|$)~\cite{dipole1} causes a negligible effective coupling interaction between the $s$ and $p$ modes at adjacent waveguides, as we confirmed experimentally below. The dynamical excitation of an orthogonal mode on a neighbour waveguide is proportional to the ratio $V_{sp}/\Delta \beta$, where $V_{sp}$ is the NN coupling interaction between the s and p modes. For standard elliptical waveguides~\cite{dipole1}, $V_{sp}/\Delta \beta\sim1/30$. (In atomic systems, this is related to 
 the energy difference between different energy levels. Note that the coupling interaction between the $s$ and $p$ modes on adjacent sites can induce interesting phenomena, such as topological edge modes~\cite{li1}; however, its experimental atomic implementation is still a challenge.) By following these considerations, now we can write the dynamical equations for both modes just by identifying $j=s$ or $p$ in Eq.~(\ref{eq1}) and by writing $V_i$ as $V_i^j$, to distinguish the coupling constants for different modes (in general, as the wave-function of the fundamental mode has a shorter evanescent tail~\cite{li1,dipole1}, $|V_i^p|> |V_i^s|$). To simplify the description, we will consider only NN coupling such that $V_1,V_2\gg V_3,V_4$, and a detuning $\Delta\beta\equiv\beta_s-\beta_p\approx30$ cm$^{-1}$~\cite{dipole1}. In Fig.~\ref{f2}, we present an example of the composed linear spectrum for this two-mode-system. We observe four flat bands located {\color{SMblack}at} $\pm V_1^p$ and $\Delta\beta\pm V_1^s$, and also the corresponding FB mode profiles. These states satisfy a destructive interference condition at connector sites (white zero amplitudes at the central row), depending on the sign of coupling constants. The relative sign of the coupling coefficients is determined by the parity symmetry of the $s$ and $p$ modes, considering the profiles sketched in Figs.~\ref{f2}. Whereas the fundamental coupling constants are always positive ($V_1^s,V_2^s>0$), the dipolar ones are determined by the specific geometry: $V_1^p>0$ and $V_2^p<0$. In Fig.~\ref{f2} we observe that the simplest fundamental FB mode ($a$) possesses the larger longitudinal propagation constant $k_z$, while the more complex dipolar one ($d$) has a shorter value, for this two-modes system.

\begin{figure}[t!]
\centering
\includegraphics[width=0.8\linewidth]{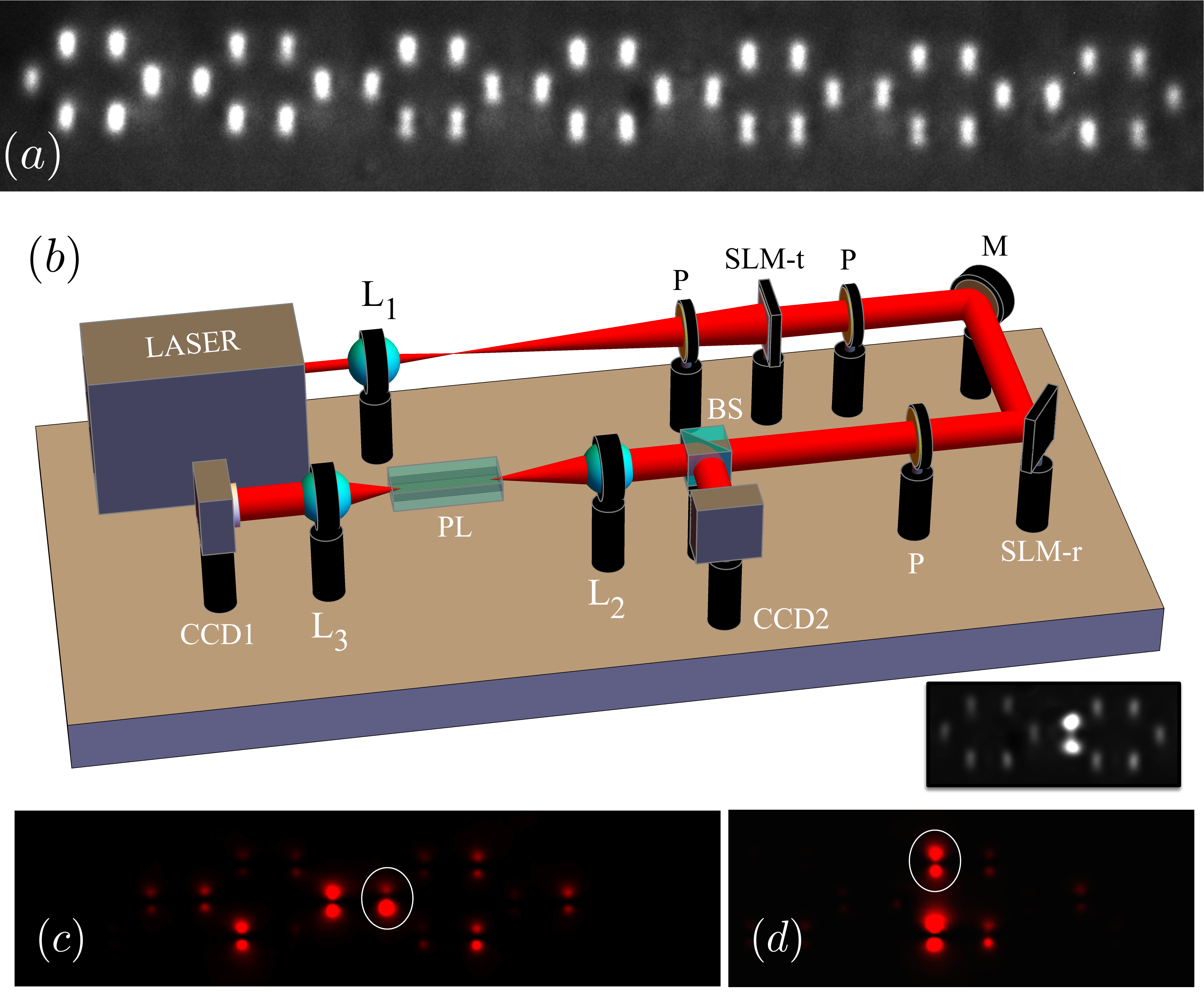}
\caption{{\bf } (a) White-light-micrograph of the output facet of a graphene ribbon (PL). (b) Experimental setup to excite and measure the dynamics of a given input state. Here, L: lens; P: polariser; SLM-t (SLM-r): transmission (reflection) SLM; M: Mirror; BS: Beam Splitter; CCD: Camera; PL: Photonic Lattice. The inset shows a dipolar input state launched at an $A$ site. (c) and (d) show the output intensity profiles of the dipolar single-site excitations injected separately at the $A$ and $B$ sites, respectively, as indicated with the white circles. To highlight the low intensities, a nonlinear colour-map was used.
\label{f3}}
\end{figure}


{\it Experiments.} Photonic graphene ribbons are directly fabricated inside a borosilicate substrate (Corning Eagle$^{2000}$) using ultrafast laser inscription~\cite{ol96, ol17sm}. Our fabrication method produces waveguides which are elongated along the vertical direction, therefore, the dipolar ($p$) modes are constrained to exist in that direction too. In Fig.~\ref{f3}(a), a white-light transmission micrograph of the output facet is presented, showing the vertically oriented waveguides. The laser-writing parameters are optimised to produce single-mode waveguides with low propagation losses at a $780$~nm wavelength. The final lattices are inscribed in a $3$-cm-long substrate, with $a\!=\!17\ \mu$m waveguide spacing. In order to study the dynamics of the $s$ and $p$ modes, we reduced the wavelength to perform the experiment at $\lambda=640$ nm. We implement an \textit{image generator setup}~\cite{dipole1} as shown in Fig.~\ref{f3}(b), which enables us to generate an arbitrary input state that can be launched on the photonic lattice PL (this is mounted on a $5$-axis-stage, which is not shown in the figure). The key element of this setup is a sequence of two spatial light modulators (SLM), that modulate the amplitude (SLM-t) and the phase (SLM-r) of an incident laser beam. Using this configuration, we launch a desired input state (with specific spatial profile and intensity and/or phase distribution) at a given lattice site. For example, the inset in Fig~\ref{f3}(b) shows a dipolar input state generated by the image generator setup. Figs.~\ref{f3}(c,d) present the output intensity distributions for single-site dipolar excitations at $A$ and $B$ sites, respectively. We observe how the energy diffracts in the lattice, due to the excitation of dispersive bands in the spectrum.

\begin{figure}[t!]
\centering
\includegraphics[width=0.8\linewidth]{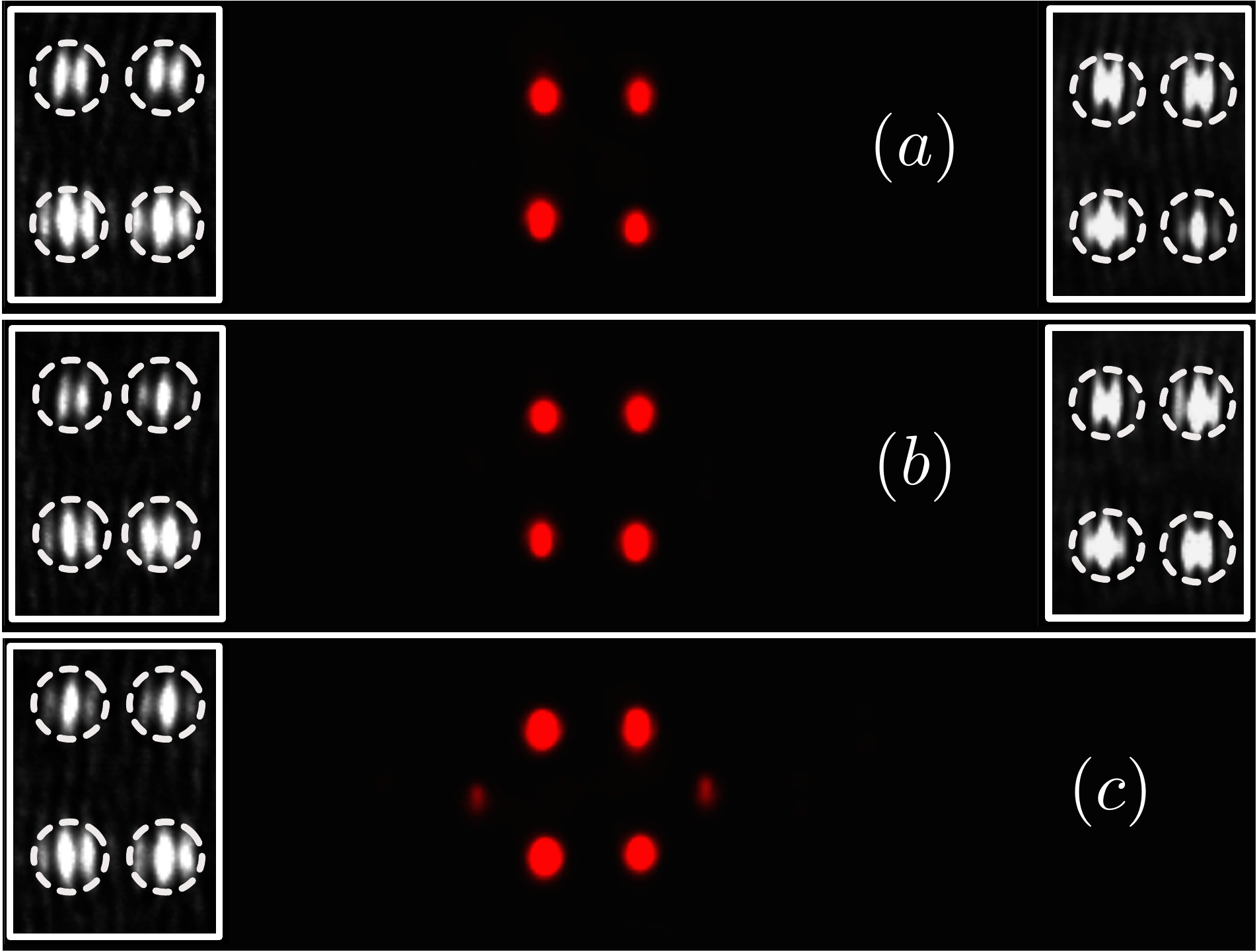}
\caption{{\bf } Output profiles for different input conditions: fundamental FB profiles at (a) $k_z=\Delta\beta+V_1^s$ and (b) $k_z=\Delta\beta-V_1^s$, and (c) four in-phase sites. Left (right)-insets: interferogram of a tilted plane wave with input (output) profiles. To highlight the low intensities, a nonlinear colour-map was used.
\label{f4}}
\end{figure}

In order to observe the dynamics of flat-band states, we use our image generator setup to excite only four desired sites of a unit cell with different spatial and phase profiles. First, to excite the fundamental FB modes, we generate two input states as sketched in Figs.~\ref{f2}(a) and (b). The observed outputs are presented in Figs.~\ref{f4}(a) and (b), respectively. We see that both FB input states propagate along the crystal without exhibiting any significant diffraction across the lattice, with an evident zero background. These states remain quite localised in space and occupy only four sites of the lattice, constituting two completely independent orthogonal states. To measure the phase profile of the input as well as of the output states, we implement an interferogram setup [this is not shown in Fig.~\ref{f3}(b) and simply consists on superposing the output profile with an extended tilted plane wave]. The left and right insets in Figs.~\ref{f4}(a) and (b) show the input and output phase structure, respectively. As the intensity and phase profiles are preserved in the dynamics, we can confirm the first excitation of the two fundamental FB modes. Additionally, we inject an in-phase four-sites excitation pattern and observe that the energy starts to spread to the rest of the lattice by the excitation of $A$ and $D$ connector sites [see Fig.~\ref{f4}(c)]. This input condition {\color{SMblack}excites} most of the linear spectrum and, therefore, for a longer propagation distance or a shorter waveguide separation, the energy would spread faster and would cover a larger transverse area, as we have confirmed numerically.

\begin{figure}[t!]
\centering
\includegraphics[width=0.8\linewidth]{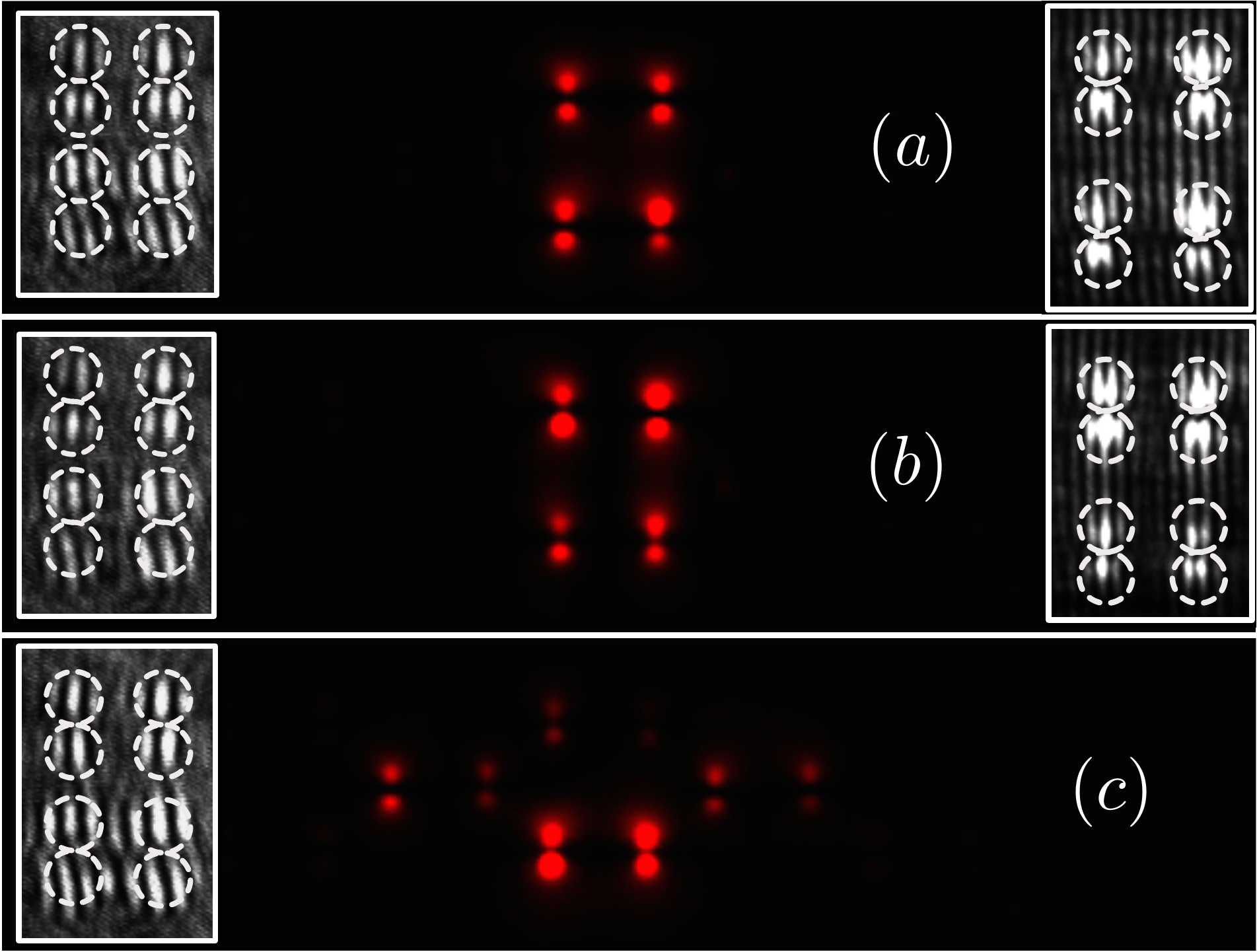}
\caption{{\bf } Output profiles for different input conditions: dipolar FB profiles at (a) $k_z=V_1^p$ and (b) $k_z=-V_1^p$, and (c) four in-phase sites. Left (right)-insets: interferogram of a tilted plane wave with input (output) profiles. To highlight the low intensities, a nonlinear colour-map was used.
\label{f5}}
\end{figure}

In the next step, we excite the dipolar ($p$) flat-band modes, which is {\color{SMblack}considerably} more challenging due to the complexity of the required spatial and phase profiles. [Note that the single-site excitations of the dipolar mode 
{\color{SMblack}were} presented in Fig.~\ref{f3}(c,d).] Precise control of the input {\color{SMblack}state,} as well as its accurate overlap with the dipolar modes of the lattice {\color{SMblack}sites (waveguides),} is required. We generated two dipolar FB modes sketched in Figs.~\ref{f2}(c) and (d) and measured outputs are shown in Figs.~\ref{f5}(a) and (b). In both cases, we observe a spatially localised state which occupies only four sites of the lattice, with a zero background. The interferograms show that the input and output phase profiles are preserved during the propagation, confirming the excitation of p-FB modes. We probe the relevance of the phase structure, on the cancellation of the transport through connector sites~\cite{luis}, by injecting an input pattern composed of four in-phase dipolar waveguide modes. In Fig.~\ref{f5}(c) we show a complete destruction of the input profile, as a consequence of exciting the dispersive part of the spectrum.

{\it Discussions}
We studied a graphene-ribbon lattice and showed the existence of $s$ and $p$ flat-band modes in the linear optical regime. Due to the symmetry of this lattice geometry, FB states can exist even in the presence of next NN interactions, as we have theoretically demonstrated by inspecting the linear properties of this system. Our lattice model possesses two FB per mode which correspond to bulk FB states, something that is particularly different to the already predicted FB edge modes in graphene-like lattices~\cite{gr2}. In our homogeneous lattice, fundamental and dipolar modes are effectively decoupled, showing no interaction between these modes. We carefully prepared several input states and experimentally observed a stable propagation of the four FB modes, what is confirmed by the analysis of the corresponding phase structure. This is the first experimental evidence of a controlled excitation of a system possessing two flat-bands per mode and, also, this is the first observation ever of a p-FB mode in any physical system. The ability to precisely control the input states in photonic lattices gives us a unique access to investigate more complex phenomena, as it has been suggested in different areas of physics~\cite{rep1,rep2,Garanovich2012,Bloch2005,Jaksch2005,li1,yin,li2}. 
It should also be highlighted that the effective coupling between the spatial modes (orbitals) can be controlled by tuning their energy (propagation constant) mismatch.  Experimental realisation of such engineered photonic lattices with interacting spatial modes will enable us to investigate intriguing phenomena~\cite{yin,li1} with more complex dynamics.

{\it Fabrication.} The photonic devices, used in the experiments, were fabricated using the ultrafast laser inscription~\cite{ol96}. By focusing sub-picosecond laser pulses (Menlo System BlueCut) inside a borosilicate substrate (Corning Eagle$^{2000}$), the refractive index of the substrate is modified permanently.  Each optical waveguide is fabricated by translating the substrate (at 8~mm/s speed) once through the focus of the laser beam.

{\it Image generation.} We generate specific input conditions by using an \textit{image generator setup}. This setup consists of a sequence of several optical elements described as follows: We first expand a laser beam in order to cover completely the screen of a transmission spatial light modulator (SLM). By setting two polarisers, we optimize the amplitude modulation response of this SLM and create a given light pattern. In our experiments, this pattern consists 
of several light spots located on specific positions depending on the studied lattice. Then, we modulate only in phase this pattern using a reflective SLM. After this point, we obtain a light pattern which is already modulated in amplitude and phase. Finally, by using different optics and a $10\times$ microscope objective (MO), we decrease the size of the pattern in order to match the specific dimensions of the lattice. To calibrate this, we install a beam splitter before the MO and take an image on a CCD camera using the reflected light on this facet. By launching white light on the output facet, we are able to observe the waveguide positions and check the dimensions of the generated image in the input facet. Finally, we take several images of the output patterns by installing after the sample another $10\times$ MO and a second CCD camera. \\

{\it Acknowledgment.} The authors sincerely thank financial support from Programa ICM RC130001, FONDECYT Grant No. 1151444, UK Science and Technology Facilities Council (STFC) through ST/N000625/1.



\begin{thebibliography}{99}
%
\bibitem{gr1} K. Nakada, M. Fujita, G. Dresselhaus, and M.S. Dresselhaus, Edge state in graphene ribbons: Nanometer size effect and edge shape dependence. Phys.~Rev. B \textbf{54}, 17954 \href{https://doi.org/10.1103/PhysRevB.54.17954}{(1996)}.
%
\bibitem{gr2}Y.-W. Son, M. L. Cohen, and S. G. Louie, Energy Gaps in Graphene Nanoribbons. Phys.~Rev.~Lett. \textbf{97}, 216803 \href{https://doi.org/10.1103/PhysRevLett.97.216803}{(2006)}.
%
\bibitem{gr3}M.Y. Han, B. \"Ozyilmaz, Y. Zhang, and P. Kim, Energy Band-Gap Engineering of Graphene Nanoribbons. Phys.~Rev.~Lett. \textbf{98}, 206805 \href{https://doi.org/10.1103/PhysRevLett.98.206805}{(2007)}.
%
\bibitem{gr4}L. Tapaszt\'o, G. Dobrik, P. Lambin, and L.P. Bir\'o, Tailoring the atomic structure of graphene nanoribbons by scanning tunnelling microscope lithography. Nature Nanotech. \textbf{3}, 397 \href{https://doi.org/10.1038/nnano.2008.149}{(2008)}.
%
\bibitem{gr5}D.V. Kosynkin, A.L. Higginbotham, A. Sinitskii, J.R. Lomeda, A. Dimiev, B. Katherine Price, and J.M. Tour, Longitudinal unzipping of carbon nanotubes to form graphene nanoribbons. Nature \textbf{458}, 872-876 \href{https://doi.org/10.1038/nature07872}{(2009)}.
%
\bibitem{gr6}J. Baringhaus, M. Ruan, F. Edler, A. Tejeda, M. Sicot, A. Taleb-Ibrahimi, A.-P. Li, Z. Jiang, E.H. Conrad, C. Berger, C. Tegenkamp, and W.A. de Heer, Exceptional ballistic transport in epitaxial graphene nanoribbons. Nature \textbf{506}, 349 \href{https://doi.org/10.1038/nature12952}{(2014)}.
%
\bibitem{gr7}A. Kimouche, M.M. Ervasti, R. Drost, S. Halonen, A. Harju, P.M. Joensuu, J. Sainio, and P. Liljeroth, Ultra-narrow metallic armchair graphene nanoribbons. Nat. Commun. \textbf{6}, 10177 \href{https://doi.org/10.1038/ncomms10177}{(2015)}.
%
\bibitem{gr8}F. Bonaccorso, Z. Sun, T. Hasan, and A. C. Ferrari, Graphene photonics and optoelectronics. Nat. Photon. \textbf{4}, 611 \href{https://doi.org/10.1038/nphoton.2010.186}{(2010)}.
%
\bibitem{hcmoti}O. Peleg, G. Bartal, B. Freedman, O. Manela, M. Segev, and D.N. Christodoulides, Conical Diffraction and Gap Solitons in Honeycomb Photonic Lattices. Phys.~Rev.~Lett. \textbf{98}, 103901 \href{https://doi.org/10.1103/PhysRevLett.98.103901}{(2007)}.
%
\bibitem{hcedge}Y. Plotnik, M.C. Rechtsman, D. Song, M. Heinrich, J.M. Zeuner, S. Nolte, Y. Lumer, N. Malkova, J. Xu, A. Szameit, Z. Chen, and M. Segev, Observation of unconventional edge states in `photonic graphene'. Nature Mater. \textbf{13}, 57 \href{https://doi.org/10.1038/nmat3783}{(2013)}.
%
\bibitem{htopo}M.C. Rechtsman, J.M. Zeuner, Y. Plotnik, Y. Lumer, D. Podolsky, F. Dreisow, S. Nolte, M. Segev, and A. Szameit, Photonic Floquet topological insulators. Nature \textbf{496}, 196 \href{https://doi.org/10.1038/nature12066}{(2013)}.
%
\bibitem{hcpseudo}D. Song, V. Paltoglou, S. Liu, Y. Zhu, D. Gallardo, L. Tang, J. Xu, M. Ablowitz, N.K. Efremidis, and Z. Chen, Unveiling pseudospin and angular momentum in photonic graphene. Nat. Commun. \textbf{6}, 6272 \href{https://doi.org/10.1038/ncomms7272}{(2015)}.

%
\bibitem{rep1}F. Lederer, G.I. Stegeman, D.N. Christodoulides, G. Assanto, M. Segev, and Y. Silberberg, Discrete solitons in optics. Phys. Rep. \textbf{463}, 1 \href{https://doi.org/10.1016/j.physrep.2008.04.004}{(2008)}.
%
\bibitem{rep2}S. Flach and A. Gorbach, Discrete breathers-- Advances in theory and applications. Phys. Rep. {\bf 467}, 1 \href{https://doi.org/10.1016/j.physrep.2008.05.002}{(2008)}.
\bibitem{Garanovich2012}I. L. Garanovich, S. Longhi, A. A. Sukhorukov, and Y. S. Kivshar, Light propagation and localization in modulated photonic lattices and waveguides. Phys. Rep. {\bf 518,} 1 \href{https://doi.org/10.1016/j.physrep.2012.03.005}{(2012)}.
%
\bibitem{Bloch2005}I. Bloch, Ultracold quantum gases in optical lattices. Nat. Phys. {\bf1,} 23 \href{https://doi.org/10.1038/nphys138}{(2005)}.
%
\bibitem{Jaksch2005}D. Jaksch and P. Zoller, The cold atom Hubbard toolbox. Ann. Phys. {\bf315,} 52 \href{https://doi.org/10.1016/j.aop.2004.09.010}{(2005)}.
\bibitem{Dreisow2008}F. Dreisow, M. Heinrich, A. Szameit, S. Doering, S. Nolte, A. T{\"u}nnermann, S. Fahr, and F. Lederer, Spectral resolved dynamic localization in curved fs laser written waveguide arrays. Opt. Express {\bf16,} 3474 \href{https://doi.org/10.1364/OE.16.003474}{(2008)}.
\bibitem{WSL}S. Mukherjee, A. Spracklen, D. Choudhury, N. Goldman, P. \"Ohberg, E. Andersson, and R. R. Thomson, Modulation-assisted tunneling in laser-fabricated photonic Wannier-Stark ladders. New J. Phys. {\bf17,} 115002 \href{https://doi.org/10.1088/1367-2630/17/11/115002}{(2015)}.

\bibitem{Schwartz2007}T. Schwartz, G. Bartal, S. Fishman, and M. Segev, Transport and Anderson localization in disordered two-dimensional photonic lattices. Nature {\bf 446,} 52 \href{https://doi.org/10.1038/nature05623}{(2007)}.
%
\bibitem{Billy2008}J. Billy, V. Josse, Z. Zuo, A. Bernard, B. Hambrecht, P. Lugan, D. Cl\'ement, L. Sanchez-Palencia, P. Bouyer, and A. Aspect, Direct observation of Anderson localization of matter waves in a controlled disorder. Nature (London) {\bf 453,} 891 \href{https://doi.org/10.1038/nature07000}{(2008)}.
\bibitem{Greiner2002}M. Greiner, O. Mandel, T. Esslinger, T.W. H\"ansch and I.Bloch, Quantum phase transition from a superfluid to a Mott insulator in a gas of ultracold atoms. Nature {\bf 415,} 39 \href{https://doi.org/10.1038/415039a}{(2002)}.
%
\bibitem{Szameit2006}A. Szameit, J. Burghoff, T. Pertsch, S. Nolte, A. T\"unnermann, and F. Lederer, Two-dimensional soliton in cubic fs laser written waveguide arrays in fused silica. Opt. Express {\bf14,} 6055 \href{https://doi.org/10.1364/OE.14.006055}{(2006)}.
\bibitem{fslt}A. Szameit and S. Nolte S, Discrete optics in femtosecond-laser-written photonic structures. J. Phys. B: At. Mol. Opt. Phys. \textbf{43}, 163001 \href{https://doi.org/10.1088/0953-4075/43/16/163001}{(2010)}.
%
\bibitem{femtoseba}Y. Bellouard, A. Champion, B. McMillen, S. Mukherjee, R. R. Thomson, C. P\'epin, P. Gillet, and Y. Cheng, Stress-state manipulation in fused silica via femtosecond laser irradiation. Optica {\bf 3}, 1285-1293 \href{https://doi.org/10.1364/OPTICA.3.001285}{(2016)}.
%
\bibitem{lieb2}R.A. Vicencio, C. Cantillano, L. Morales-Inostroza, B. Real, C. Mej\'ia-Cort\'es, S. Weimann, A. Szameit, and M. I. Molina, Observation of Localized States in Lieb Photonic Lattices. Phys.~Rev.~Lett. \textbf{114}, 245503 \href{https://doi.org/10.1103/PhysRevLett.114.245503}{(2015)}.
%
\bibitem{liebseba}S. Mukherjee, A. Spracklen, D. Choudhury, N. Goldman, P. {\"O}hberg, E. Andersson, and R. R. Thomson, Observation of a Localized Flat-Band State in a Photonic Lieb Lattice. Phys.~Rev.~Lett. \textbf{114}, 245504 \href{https://doi.org/10.1103/PhysRevLett.114.245504}{(2015)}.
%
\bibitem{bic}J. von Neumann and E. Wigner, {\"U}ber merkw\"urdige diskrete Eigenwerte. Phys. Z. \textbf{30}, 465 \href{https://doi.org/10.1007/978-3-662-02781-3_19}{(1929)}.
%
\bibitem{desa1}D. Leykam, O. Bahat-Treidel, and A.S. Desyatnikov, Pseudospin and nonlinear conical diffraction in Lieb lattices. Phys.~Rev.~A \textbf{86}, 031805(R) \href{https://doi.org/10.1103/PhysRevA.86.031805}{(2012)}.
%
\bibitem{luis}L. Morales-Inostroza and R.A. Vicencio, Simple method to construct flat-band lattices. Phys.~Rev.~A \textbf{94}, 043831 \href{https://doi.org/10.1103/PhysRevA.94.043831}{(2016)}.
\bibitem{becdipole}G. Wirth, M. \"Olschl\"ager, and A. Hemmerich, Evidence for orbital superfluidity in the $P$-band of a bipartite optical square lattice. Nature Physics \textbf{7}, 147 \href{https://doi.org/10.1038/nphys1857}{(2011)}.
%
\bibitem{dipole1}C. Cantillano, L. Morales-Inostroza, B. Real, S. Rojas-Rojas, A. Delgado, A. Szameit, and R.A. Vicencio, Observation of dipolar transport in one-dimensional photonic lattices. Science Bulletin \textbf{62} (5), 339-344 \href{https://doi.org/10.1016/j.scib.2017.01.035}{(2017)}.
%
\bibitem{wu}C. Wu, D. Bergman, L. Balents, and S. Das Sarma, Flat Bands and Wigner Crystallization in the Honeycomb Optical Lattice. Phys.~Rev.~Lett. \textbf{99}, 070401 \href{https://doi.org/10.1103/PhysRevLett.99.070401}{(2007)}.
%
\bibitem{li1}X. Li, E. Zhao, and W.V. Liu, Topological states in a ladder--like optical lattice containing ultracold atoms in higher orbital bands. Nat. Commun. \textbf{4}, 1523 \href{https://doi.org/10.1038/ncomms2523}{(2013)}.
%
\bibitem{yin}S. Yin, J. E. Baarsma, M. O. J. Heikkinen, J. P. Martikainen, and P. T\"orma, Superfluid phases of fermions with hybridized $s$ and $p$ orbitals. Phys.~Rev.~A \textbf{92}, 053616 \href{https://doi.org/10.1103/PhysRevA.92.053616}{(2015)}.
%
\bibitem{li2}X. Li and W.V. Liu, Physics of higher orbital bands in optical lattices: a review. Rep. Prog. Phys. \textbf{79}, 116401 \href{https://doi.org/10.1088/0034-4885/79/11/116401}{(2016)}.
%
\bibitem{ol96} K. M. Davis, K. Miura, N. Sugimoto, and K. Hirao, Writing waveguides in glass with a femtosecond laser. Opt. Lett. \textbf{21}, 1729 \href{https://doi.org/10.1364/OL.21.001729}{(1996)}.
%
\bibitem{ol17sm} S. Mukherjee and R. R. Thomson, Observation of robust flat-band localization in driven photonic rhombic lattices. Opt. Lett. \textbf{42} (12), 2243-2246 \href{https://doi.org/10.1364/OL.42.002243}{(2017)}.
\end{thebibliography}
\end{document}